\documentclass[twocolumn,showpacs,preprintnumbers,amsmath,amssymb]{revtex4}

\usepackage{graphicx}
\usepackage{dcolumn}
\usepackage{bm}
\usepackage{amsmath,epsfig}

\begin{document}



\title{A tool for filtering information in complex systems}

\author{M. Tumminello $^{(1)}$, T. Aste $^{(2)}$, T. Di Matteo $^{(2)}$, R.N. Mantegna $^{(1,3)}$ }

\affiliation{$^{(1)}$ 
INFM Unit\`a di Palermo and Dipartimento di Fisica e Tecnologie Relative, 
Universit\`{a} di Palermo, Viale delle Scienze, Palermo, I-90128, Italy\\
$^{(2)}$ Department of  Applied Mathematics, Australian National University
0200 Canberra, ACT, Australia\\
$^{(3)}$ INFN, Sezione di Catania, Catania, Italy}

\date{\today}

\begin{abstract}
We introduce a technique to filter out complex data-sets by extracting a subgraph of representative links. 
Such a filtering can be tuned up to any desired level by controlling the genus of the resulting graph. 
We show that this technique is especially suitable for correlation based graphs giving filtered graphs which preserve the hierarchical organization of the minimum spanning tree but containing a larger amount of information in their internal structure.
In particular in the case of planar filtered graphs (genus equal to 0) triangular loops and 4 element cliques are formed.
The application of this filtering procedure to 100 stocks in the USA equity markets shows that such loops and cliques have important and significant relations with the market structure and properties.\\
\end{abstract}

\maketitle

\section{Introduction}
Several complex systems have been recently investigated from the perspective of the (weighted) networks that are linking the different elements comprising them \cite{Strogatz,Albert2002,Dorogovtsev,Newman2003}. 
Indeed, complex systems are in general made of \emph{several interacting elements} and it is rather natural to associate to each element a node and to each interaction a link yielding to a graph. 
Examples include food webs \cite{Garlaschelli}, scientific citations 
\cite{Redner98}, social networks \cite{Wasserman,Newman2002}, communication networks \cite{Pastor2001}, sexual contacts among individuals \cite{Liljeros}, company links in a stock portfolio \cite{Mantegna99}, the Internet \cite{Faloutsos1999} and the World Wide Web \cite{Albert99}. 
The properties of such graphs have been studied with the aim of catching basic features of the investigated systems \cite{Watts98,Barabasi99,Amaral00}.
However, the complexity of the system is in general reflected in the associated graph which results in an intricate interweaved and densely connected structure. 
There is therefore a general need to find methods which are able to single out the key information by filtering such a complex graph into a simpler relevant subgraph. 
Such a filtering is especially essential for correlation-based graphs where, in the absence of any filtering procedure, all links among elements are present. \\
In this paper we introduce a new filtering procedure which extracts a representative sub graph with a controlled complexity  and maximal information content out of the graph describing the system. 
To illustrate the method we present a concrete example dealing with 100 stocks belonging to a USA equity portfolio. 
In the modeling of equity portfolios a natural starting point is the investigation of cross-correlation among time series of returns of stock pairs. The correlation provides a similarity measure among the behavior of different elements in the system. It was shown by one of the authors that a powerful method to investigate financial systems consists in the extraction of a minimal set of relevant interactions associated with the strongest correlations belonging to the Minimum Spanning Tree (MST) \cite{Mantegna99}. However, the reduction to a minimal skeleton of links is necessarily very drastic in filtering correlation based networks loosing therefore valuable information. The necessity of a less drastic filtering procedure has been already raised in the literature. 
For example, an extension from trees to more general graphs generated by selecting the most correlated links has been proposed in Refs. \cite{Onnela03}. 
However, with the method discussed in Refs. \cite{Onnela03} is highly improbable to obtain a  filtered network connecting all elements via some path by retaining a number of links of the same order of the number of elements.\\
The method that we present in this paper is based on the key-idea that graphs with different degrees of complexity can be constructed by iteratively linking the most strongly connected nodes under the constraint of generating graphs that can be embedded on a surface of a given genus $g=k$  \cite{Noi}. The genus is a topologically invariant property of a surface defined as the largest number of non-isotopic simple closed curves that can be drawn on the surface without separating it, i.e. the number of handles in the surface.
We prove that such graphs have the same hierarchical tree associated to the MST \cite{Gower, Mantbook} but contain a larger amount of information which increases with the genus. 
We show that, with respect to the MST, the major relative improvement of the information stored in the graph is realized for the planar case when the genus assumes the value $k=0$.\\

\section{The filtering procedure}
\textbf{Construction algorithm.} Let us first illustrate the method and the associated algorithm to filter significant information out of a given complex system composed by $n$ elements where a similarity measure $S$ between pairs of elements is defined, e.g. the weight of links in the original network or the correlation coefficient matrix of the system. 
An ordered list $S_{ord}$ of pair of nodes can be constructed  by arranging them in a descending order accordingly with the value of the similarity $s_{ij}$ between element $i$  and element $j$. Let us first consider the construction algorithm for the MST:  {\it following the ordered list  $S_{ord}$ starting from the couple of elements with larger similarity one adds an edge between element $i$  and element $j$ if and only if the graph obtained after the edge insertion is still a forest or it is a tree}. A forest is a disconnected graph in which any two elements are connected by at most one path, i.e. a disconnected ensemble of trees.
With this procedure the graph obtained after all links of $S_{ord}$ are considered is the MST. In fact when the last link is included in the graph the forest reduces to a tree.\\
In direct analogy with this construction of  the MST, we construct graphs by connecting elements with largest similarity under the topological constraint of fixed genus $g=k$. 
The construction algorithm for such graphs is:
 {\it following the ordered list $S_{ord}$ starting from the couple of elements with larger similarity one adds an edge between element $i$ and element $j$  if and only if the resulting graph can still be embedded on a surface of genus $g\leq k$ after such edge insertion}.
This generates \emph{simple}, \emph{undirected}, \emph{connected} graphs embedded on  a surface of genus $g=k$.
In the next section we demonstrate that these graphs have the same hierarchical tree of the MST and that they possess relevant additional information associated with the structure of loops and cliques making them natural extensions of the MST.
A clique of $r$ elements ($r$-clique) is a complete subgraph that links all $r$ elements. \\
A special case is when $g=0$  and the resulting graph is planar \cite{Planar}, i.e. it can be embedded on the sphere. 
This graph is the first extension of the MST and we name it Planar Maximally Filtered Graph (PMFG). An implementation of the algorithm providing the PMFG written in Mathematica is accessible as supplementary information.
A basic difference of the PMFG with respect to the MST is the number of links which is $n-1$ in the MST and $3\,(n-2)$ in the PMFG. 
On the other hand, in general, the number of links in a graph G with a fixed genus $g=k$ is at most $3 (n-2+2 k)$. 
Therefore, in most practical cases, when $k \sim O(1)$ and $n \gg 1$, the relative increase in the number of links that might be included in the graph by increasing its genus is very small. 
It follows that the PMFG assumes a special status among all the graphs constructed with the introduced algorithm. 
Indeed, it is the simplest and the one providing the most significant additional information with respect to the MST. 
For this reason we will deserve a special attention to it. It is worth noting that the construction algorithm and the topological constraints on the PMFG force each element to participate to at least a clique of 3 elements. In other words, the PMFG is a topological triangulation of the sphere. Only cliques of 3 and 4 elements are permitted in the PMFG. Indeed Kuratowski's theorem \cite{Planar} does not allow cliques with a number of elements larger than 4 in a planar graph. 
Larger cliques can only be present in graphs with  genus $k>0$. 
The larger the value of $k$ the larger is the number of elements $r$ of the maximal allowed clique (specifically $r \le \frac{7 + \sqrt{1+48 k}}{2}$ \cite{Ringel}).\\ 

\textbf{Hierarchical organization.}
We prove the following statement: {\it At any step of construction of the MST and graph G  of genus $g=k$ if two elements are connected via at least one path in one of the considered graphs then they are connected also in the other one}.
To this end we must recall the concept of bridge: a link between two elements is a bridge whenever the elements are disconnected via any path in its absence. It follows from the definition of MST that all links in the MST are bridges. On the other hand, for graphs with a fixed genus  we have the following important property: if a bridge is inserted between two previously unconnected regions of a graph G, characterized by the genus $g=k$, then the genus of the graph obtained after the insertion is still $k$. This property is straightforwardly proved as a corollary of the Miller theorem \cite{Miller} by noting that the addition of a bridge to a graph leaves unchanged the biconnected components of the graph. 
The above property implies that if the construction algorithm of G selects a link which is a bridge for the graph at that step of construction, then the link is always added to the graph. \\
We now prove the above statement by induction. In the following we indicate as $MST_m$ and $G_m$ the graphs constructed by using the similarity measure up to the $m-th$ row of $S_{ord}$. For the first two steps of construction the statement is true: $MST_2$ and $G_{2}$ graphs are always equal.  Now suppose the statement is true at the step \emph{m} of construction, i.e. for $G_m$ and $MST_m$. For the step \emph{m+1} only four cases are possible: 

(i) the new link, connecting the vertices $i$ and $j$, is a bridge for the $MST_{m+1}$. By the definition of bridge this implies that the vertices $i$ and $j$ are not connected via any path in $MST_{m}$. Therefore, by inductive hypothesis, the vertices $i$ and $j$ are not connected via any path also in $G_{m}$ and then the new link is a bridge for $G_{m+1}$ too. In this case both graphs will include the considered link and then the statement is true at the step $m+1$.

(ii) the link is a bridge for $G_{m+1}$. By using the same reasoning as in (i) this implies that the same link must also be a bridge for the $MST_{m+1}$ due to the inductive hypothesis and both graphs will include the considered link and then the statement is true at the step $m+1$.

In the remaining two cases we assume the condition that the link between the vertices $i$ and $j$ is not a bridge for both $MST_{m+1}$ and $G_{m+1}$. This is a condition that can be used without loss of generality because if the link is not a bridge for $MST_{m+1}$ (or $G_{m+1}$) then one always concludes that the link is also not a bridge for $G_{m+1}$ (or $MST_{m+1}$) by following the same reasoning of case (i).

(iii) The link is not a bridge for both $MST_{m+1}$ and $G_{m+1}$ and the genus  condition $g \leq k$ fails. In this case the link is not included to any of both graphs and the statement is again true at the step $m+1$.

(iv) The link under investigation is not a bridge for both $MST_{m+1}$ and $G_{m+1}$ and the genus condition $g \leq k$ is satisfied. In this case $G_{m+1}$ includes the link and $MST_{m+1}$ does not. However due to the fact that the link added to $G_m$ is not a bridge the connectivity between pairs of elements in $MST_{m+1}$ and in $G_{m+1}$ rests unchanged in both $MST_{m}$ and $G_m$ and the statement is also verified in this last case. 

The statement is therefore true and it has an important implication: the fact that the MST is formed only by bridges implies that the MST is always contained in any graph G of genus $g=k$ generated with the construction algorithm presented above and, as a specific case, in the PMFG. 
Moreover, this statement shows even a more important fact: the formation of connected clusters of nodes in $G_m$ coincides with the formation of the same clusters in the $MST_m$.
In other words the hierarchical tree associated to graph G coincides with the one of the MST.
It is worth noting that the construction algorithm and the associated network properties also hold true in the more general case of weighted networks and non fully-connected networks. 
In other words the algorithm is general and in the case of a non-connected graph the filtered graph $G$ of genus $g=k$ will also be a non-connected graph whereas the equivalent of MST will not be a tree but a forest. \\

\textbf{An illustrative example.} We present here a simple example showing how the PMFG provides additional information with respect to the one contained in the MST and in the associated hierarchical tree. Let us consider a simple system composed by 10 elements being characterized by the MST shown in panel A of Fig. 1. Not all the information about the similarity measure is used to obtain the MST and different similarity matrices can have the same MST and hierarchical tree but end up into different PMFGs. Two quite distinct possible PMFGs with the same underlying MST are shown in panel B and C of Fig. 1. In this figure the thicker lines are identifying links belonging to both the MST and the PMFG whereas the thinner lines belong to the PMFG only. By looking just at the MST we observe in both examples of Fig. 1 B and C two groups, specifically the one formed by vertices $\alpha i$ ($i=0,...,4$), say cluster $\alpha$, and the other composed by vertices $\beta i$ ($i=0,...,4$), say cluster $\beta$. The two clusters are connected by the link between $\alpha 0$ and $\beta 0$.  The differences between the two PMFGs can be quantified by a comparison between the link structures of these two planar graphs. In the first planar graph (panel B of Fig. 1) there are 18 intra-cluster links (connecting elements $\alpha i$ to $\alpha j$ and $\beta i$ to $\beta j$) and 6 inter-cluster links (connecting elements $\alpha i$ to $\beta j$). In the second planar graph (panel C of Fig. 1) the number of intra-cluster links is 12, equal to the number of inter-cluster links.  
Another way to quantify the differences between the two PMFGs is by analyzing the clique structure of the two planar graphs. 
In the graph of panel B a total of $7$ cliques of 4 elements is found, with two of them $\{ \alpha 0,\, \alpha 1,\, \alpha 2,\, \alpha 4 \}$ and $\{ \alpha 0,\, \alpha 2,\, \alpha 3,\, \alpha 4 \}$ being characterized by elements all belonging to the cluster $\alpha$, other two having elements internal to the cluster $\beta$  $\{ \beta 0,\, \beta 1,\, \beta 2,\, \beta 4 \}$ and $\{ \beta 0,\, \beta 2,\, \beta 3,\, \beta 4 \}$ and 3 being composed of inter-cluster elements $\{ \alpha 1,\, \alpha 4,\, \beta 0,\, \beta 4 \}$, $\{ \alpha 1,\, \beta 0,\, \beta 1,\, \beta 4 \}$ and $\{ \alpha 0,\, \alpha 1,\, \alpha 4,\, \beta 0 \}$.  Looking at the cliques of 3 elements we observe 14 intra-cluster cliques and 8 inter-cluster cliques. Quite differently, in the graph of panel C we observe just 4 cliques of 4 elements, no one of them being an intra-cluster clique, and just 4 of the 20 cliques of 3 elements are intra-cluster cliques. This analysis shows that the two considered PMFGs refine in a quite different way the information associated with the (same) MST. 
This shows that PMFG provides additional information with respect to the MST and this information is primarily associated with the nature of the observed cliques.\\

\begin{figure}
\includegraphics[width = 0.48 \textwidth]{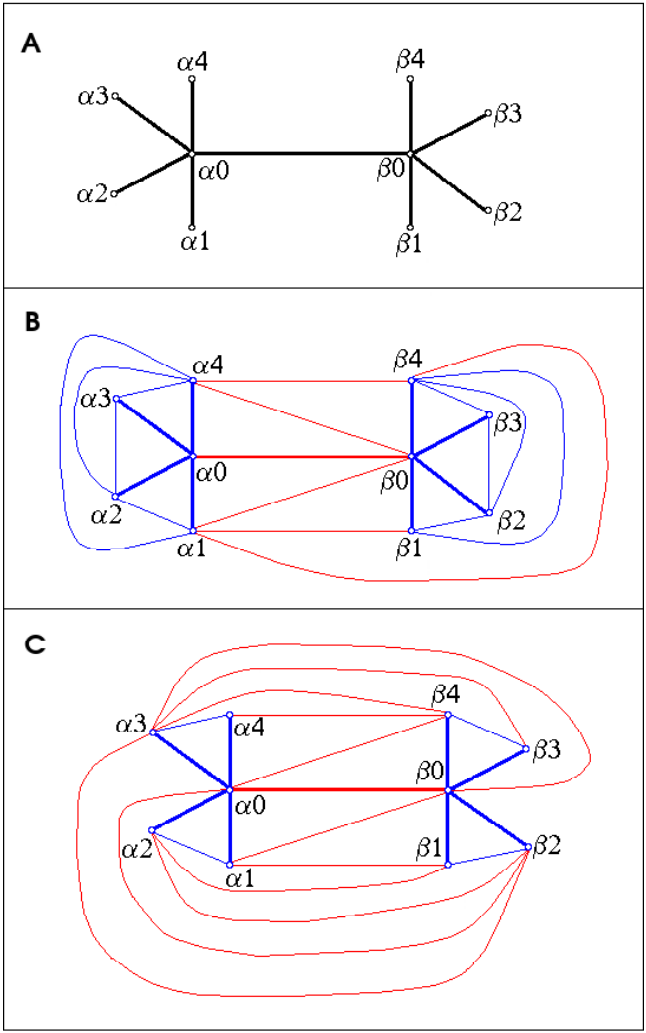}
\caption{\label{figgraphtutor} An illustrative example of two graphs that share the same MST but have distinct PMFGs. Panel A: MST of a simple system of 10 vertices. Panels B and C: PMFG of two systems with the same MST (the one drawn in panel A).  The thicker lines are identifying links belonging to both the MST and the PMFG whereas the thinner lines belong to the PMFG only.} 
\end{figure}
%

\textbf{ }
\begin{widetext}
\begin{center}
\begin{figure}[t]
\includegraphics[width = 1 \textwidth]{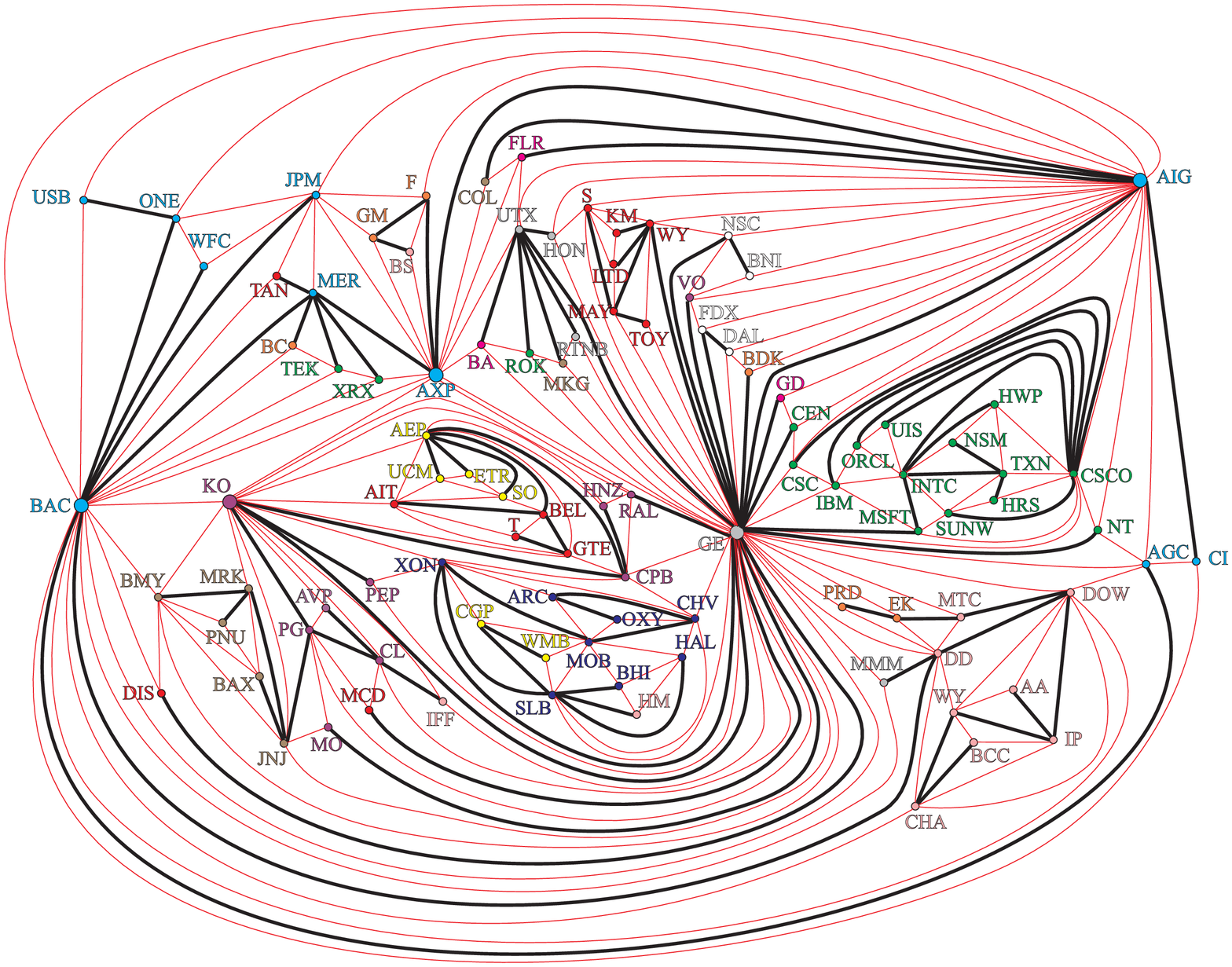}
\caption{\label{figgraph} PMFG obtained from the fully connected graph associated with the correlation coefficient matrix of 100 most capitalized stocks traded in the USA equity markets during the time period from 1995 to 1998. Cross correlation is computed by using daily returns of stocks. Stocks are indicated with their tick symbols. For information about a specific tick symbol see additional material. The graph is topologically planar: it can be drawn on the plane without edge-crossings. 
The thicker lines are belonging to the associated minimum spanning tree. It should be noted that link lengths do not reflect the value of the similarity measure between vertices.}
\end{figure}
\end{center}
\end{widetext}
%
%
\section{An empirical application} 
\textbf{PMFG for 100 US stocks.} In the previous section we have introduced a {\it general} method for constructing graphs of defined genus $g=k$ by using a similarity measure. 
In the following we constraint ourselves to the case $g=0$, i.e. to the PMFG and we present an example concerning the filtering of a graph obtained with a correlation based procedure. Specifically, we consider pair correlation between daily returns of a set of 
100 stocks traded in  the New York equity markets in the time period between 1/1995 and 12/1998 \cite{Mantegna1}. 
In this case the measure of similarity $S$ is given by the correlation coefficient $\rho_{i\,j}$ between stocks. The PMFG for this system is shown in Fig. 2. In the figure the various elements are connected through links with thicker lines indicating links belonging to both  the MST and the PMFG.
%
%
By a comparison between the PMFG and the MST several new details emerge with the most striking difference being that the PMFG allows the existence of loops and cliques as we saw in the above example.
In the PMFG we counted $292 = 3n-8$ cliques of 3 elements out of the possible $\binom{n}{3}=161,700$. 
The number of cliques of $4$ elements is $97 = n-3$. 
This number is much smaller than the number of possible cliques of $4$ elements present in the fully connected graph which are $\binom{n}{4}\cong3.92 \cdot 10^6$. 
The complete lists of cliques with 3 and 4 elements present in the PMFG are accessible as supplementary material. 
Interestingly, these numbers of 3- and 4- elements cliques coincide with the numbers of such cliques attainable when a graph is made by a set of tetrahedra  (4-cliques) packed together by sharing a triangular face.
The fact that we observe such numbers of cliques can be qualitatively explained. 
Consider $3$ elements of a correlation based network, say A, B and C. 
If A is strongly correlated to B and B is strongly correlated to C then it should also be detected a strong correlation between A and C which makes highly   probable the formation of a triangular clique.
Now, if one of these 3 elements is strongly correlated with a 4th one, say D, then also the other two are likely to have a strong correlation with D generating in this way a 4-clique: a tetrahedron.
Given the topological constraint of planarity, the next most correlated element can only be connected to maximum 3 of the 4 elements of such tetrahedron. 
The connection of a new element to three elements of the 4-clique generates another 4-clique which is a new tetrahedron sharing a face with the previous one.
By following this reasoning, we expect therefore that the basic structures in the resulting graph are the 4-cliques which during the formation of the PMFG clusterize together locally at similar correlation values and then connect to each other by following the MST as skeleton structure.
Therefore, if such $4$ elements cliques are the `building blocks' of the PMFG, then there must be strong relations between their properties and the ones of the system of 100 stocks from which they have been generated.
These relations are explored in the following section.\\

\begin{table}\nonumber
\caption{Strongest correlated intra-sector 4-cliques}
\begin{tabular}{||l|c|l|l|l|l|c|c||}
\tableline
Sec. & num & Stock 1 & Stock 2 & Stock 3 & Stock 4 & $< \rho >$ & $<y>$\\
\tableline
\tableline
E & 5 &ARC & CHV & MOB & XON & $0.628$ &  $0.335$ \\
B & 4 & BCC & CHA & IP & WY & $0.592$ &  $0.334$ \\
F & 6 & AXP & BAC & JPM & MER & $0.589$ &  $0.334$ \\
T & 8 & CSCO & INTC & MSFT & SUNW & $0.537$ &  $0.335$ \\
H & 2 & BAX & BMY & JNJ & MRK & $0.465$ &  $0.339$ \\
C & 2 & AVP & CL & KO & PG & $0.462$ &  $0.337$ \\
S & 3 & AIT & BEL & GTE & T & $0.422$ &  $0.354$ \\
U & 1 & AEP & ETR & SO & UCM & $0.398$ &  $0.343$ \\
\tableline
\tableline
\end{tabular}
\end{table}
\textbf{Financial market properties and 4-cliques structure.} Let us first classify each stock accordingly with an economic sector following the classification of the Forbes Magazine.
An analysis on all the 4-cliques in the PMFG reveals a high degree of homogeneity with respect to the economic sectors. 
Indeed, we observe that $31$ of the $97$ cliques are composed by stocks belonging to the same economic sector; 22 are composed by 3 stocks belonging to the same sector; 37 have 2 stocks from the same sector and only 7 have stocks all from different sectors.

In Table I we list the $8$ cliques with the largest mean correlation $< \rho >$ among stocks for each economic sector having at least one clique of four elements. We label the economic sectors  as Energy (E), Basic Materials (B), Financial (F), Technology (T), Healthcare (H), Consumer non cyclical (C), Services (S) and Utilities (U). The total number of intra-sector 4-cliques ($c_4$) for each sector is given in the second column of the table. It should be noticed that  $< \rho >$ among stocks is different for different sectors. For example the clique with the largest mean correlation is a clique of the Energy sector which has $<\rho>=0.628$. Whereas the clique of the sector Utilities has the smallest mean correlation with $< \rho >=0.398$. 
To better understand the structure of such cliques it is interesting to quantify how much the correlation among the stocks is spread within the clique.
In analogy to Ref. \cite{Vespignani} we compute the quantity $<y>$ inside a clique as the mean value of the \emph{disparity measure} $y (i) = \displaystyle{\sum_{j\neq i, j \in clique}} \left[\frac{\rho_{i\,j}}{s_i} \right]^2$ over the clique, where $i$ is a generic element of the clique and $s_i = \displaystyle{\sum_{j\neq i, j \in clique}}\rho_{i\,j}$ is the \emph{strength} of the element $i$.  This definition is meaningful if $\rho_{i\,j} \ge 0$ as in the case considered. 
The value of the disparity is expected to be close to $1/3$ for 4-cliques characterized by links with comparable values of the similarity measure. An inspection of the last column of Table I shows that most of the cliques have a disparity measure very close to $1/3$. Exceptions are the cliques of the sectors Services and Utilities that have a slightly smaller homogeneity in the pair correlation between stocks belonging to the cliques.

\begin{table}
\caption{4-cliques belonging to the Technology sector}
\begin{tabular}{||l|l|l|l|c|c||}
\tableline
Stock 1 & Stock 2 & Stock 3 & Stock 4 & $< \rho >$ & $<y>$\\
\tableline
\tableline
CSCO & INTC & MSFT & SUNW & $0.537$ &  $0.335$ \\ 
CSCO & IBM & INTC & MSFT & $0.534$ &  $0.335$ \\ 
CSCO & INTC & SUNW & TXN & $0.519$ &  $0.335$ \\ 
CSCO & HWP & INTC & TXN & $0.503$ &  $0.336$ \\ 
CSCO & IBM & INTC & ORCL & $0.475$ &  $0.336$ \\ 
HWP & INTC & NSM & TXN & $0.471$ &  $0.339$ \\ 
CSCO & HRS & SUNW & TXN & $0.435$ &  $0.338$ \\ 
CSCO & INTC & ORCL & UIS & $0.380$ &  $0.354$ \\ 
\tableline
\tableline
\end{tabular}
\end{table}

In Table II we present all the 8 cliques of $4$ elements observed for stocks belonging to the Technology sector. Note that also inside a single sector the level of correlation of the selected cliques may significantly vary. In fact it ranges from $< \rho >=0.380$ to $< \rho >=0.537$  showing that the PMFG is able to select cliques at different levels of correlation. The selection among all the possible cliques present in the fully connected graph is rather severe, in fact for the Technology sector we have $17$ elements and therefore the number of cliques of $4$ elements all belonging to this sector which are present in the fully connected graph is $2380$. In other words only 8 of the possible $2380$ cliques of $4$ elements of the fully connected graph are selected by the PMFG.\\
To elucidate the type and amount of information gained by extending the graph from the MST to the PMFG we focus in more detail on the intra-sector and inter-sector cliques found by the PMFG and, of course, absent in the MST. From Table III one notes that no 4-cliques are observed for the sectors Conglomerates (CO) composed by five stocks ($n_s=5$),  Consumer cyclical  (CC, $n_s=6$) and Transportation (TR, $n_s=4$) even if the number of stocks $n_s$ composing the sectors would potentially allow them. It should also be noted that the sector Capital Goods (CG, $n_s=3$) does not form a 3-clique. This observation shows that the PMFG conveys information that is not directly present in the market classification of Forbes nor in the MST. In fact, this information is associated with the clustering strength of economic sectors that have been detected by the MST.       

To quantify the degree of connection strength of elements we propose to consider the ratio between the number of 4-cliques ($c_4$) and of 3-cliques ($c_3$) present among $n_s$ stocks belonging to a given set and a normalizing quantity. These normalizing quantities are $n_s-3$ for 4-cliques and $3\,n_s-8$ for 3-cliques. Although we haven't a formal proof, our investigations suggest that these numbers are the maximal number of 4-cliques and 3-cliques respectively that can be observed in a PMFG of $n_s$ elements. 

\begin{table}\nonumber
\caption{Intra-sector connection strength}
\begin{tabular}{||c|c|c|c||}
\tableline
Sec & $n_s$ & $c_4/[n_s-3]$ & $c_3/[3\,n_s-8]$\\
\tableline
\tableline
E & 8 & $5/5=1$ & $16/16=1$ \\
F & 10 & $6/7\cong0.86$ & $20/22\cong0.91$ \\
T & 17 & $8/14\cong0.57$ & $26/43\cong0.60$ \\
B & 11 & $4/8=0.5$ & $14/25=0.56$ \\
H & 7 & $2/4=0.5$ & $7/13\cong0.54$ \\
U & 6 & $1/3\cong0.33$ & $4/10=0.40$ \\
S & 13 & $3/10=0.3$ & $12/31\cong0.39$ \\
C & 10 & $2/7\cong0.29$ & $8/22\cong0.36$ \\
CO & 5 & $0/2=0$ & $2/7\cong0.29$ \\
CC & 6 & $0/3=0$ & $0/10=0$ \\
TR & 4 & $0/1=0$ & $0/4=0$ \\
CG & 3 & $-$ & $0/1=0$ \\
\tableline
\tableline
\end{tabular}
\end{table}

In Table III the connection strength is presented for all the elements belonging to the economic sectors both for 4-cliques and 3-cliques. Table III clearly shows that the connection strength can be quite different across sets of elements. Specifically, elements belonging to some economic sectors are strongly connected within themselves whereas others are much less. Examples of strong connection are the elements of the Energy and Financial sectors whereas elements belonging to the Conglomerates,  Consumer cyclical, Transportation and Capital Goods sectors are weakly connected.
     
In Table IV we list the 7 cliques with all the 4 components belonging to different economic sectors. These 4-cliques provide bridging regions among areas of the PMFG populated by elements belonging to different economic sectors. This interpretation is supported by the fact that in these cliques some of the most connected stocks are present. In fact General Electric (GE) is present in all the 7 cliques whereas the American International Group (AIG) is present in $5$ of them.

\begin{table}
\caption{Inter-sector 4-cliques connecting 4 different sectors}
\begin{tabular}{||l|l|l|l|c|c||}
\tableline
Stock 1 & Stock 2 & Stock 3 & Stock 4 & $< \rho >$ & $<y>$\\
\tableline
\tableline
BAC	(F) & BMY	(H) 	& GE   (CO) & KO    (C) & $0.483$ &  $0.336$ \\ 
BAC	(F) & BMY	(H) 	& GE   (CO) & DIS   (S) & $0.435$ &  $0.337$ \\
AIG	(F) & GE	(CO) & NSC(TR) & WMT (S) & $0.423$ &  $0.339$ \\
AIG	(F) & GE	(CO) & NSC(TR) & VO    (C) & $0.400$ &  $0.340$ \\
AIG	(F) & GE	(CO) & FDX (TR) & VO    (C) & $0.374$ &  $0.345$ \\
AIG	(F) & BDK	(CC) & DAL (TR) & GE    (CO) & $0.360$ &  $0.346$ \\
AIG	(F) & CEN	(T) 	& GD   (CG) & GE    (CO) & $0.340$ &  $0.351$ \\
\tableline
\tableline
\end{tabular}
\end{table}

A joint reading of Table III and Table IV shows that economic sectors are not equivalent in the detected PMFG. Specifically sectors E, T and B are sectors of elements significantly connected among them but weakly interacting with stocks belonging to different economic sectors. Indeed no Energy and Basic Materials stocks and only one Technology stock (CEN) appear in Table IV. Quite differently the Financial Sector (F) has still elements strongly connected among them but it also participates to all the 7 inter-sector cliques of Table IV. In other words various sets of elements may or may not be clustered among themselves and may or may not be also connected to elements of other sectors. The PMFG is able to extract this information. Finally, elements of sectors CO and TR are very weakly clustered among them but are often present in cliques of Table IV meaning that stocks belonging to these sectors behave like bridges between sectors as F, T, H, S and C.
In summary Tables III and IV show how the PMFG is able to quantify the connection strength of elements of the graph via an analysis of the clique structure. An information which is only partially present in the MST and the hierarchical tree. 

\begin{figure}
\includegraphics[width = 0.48 \textwidth]{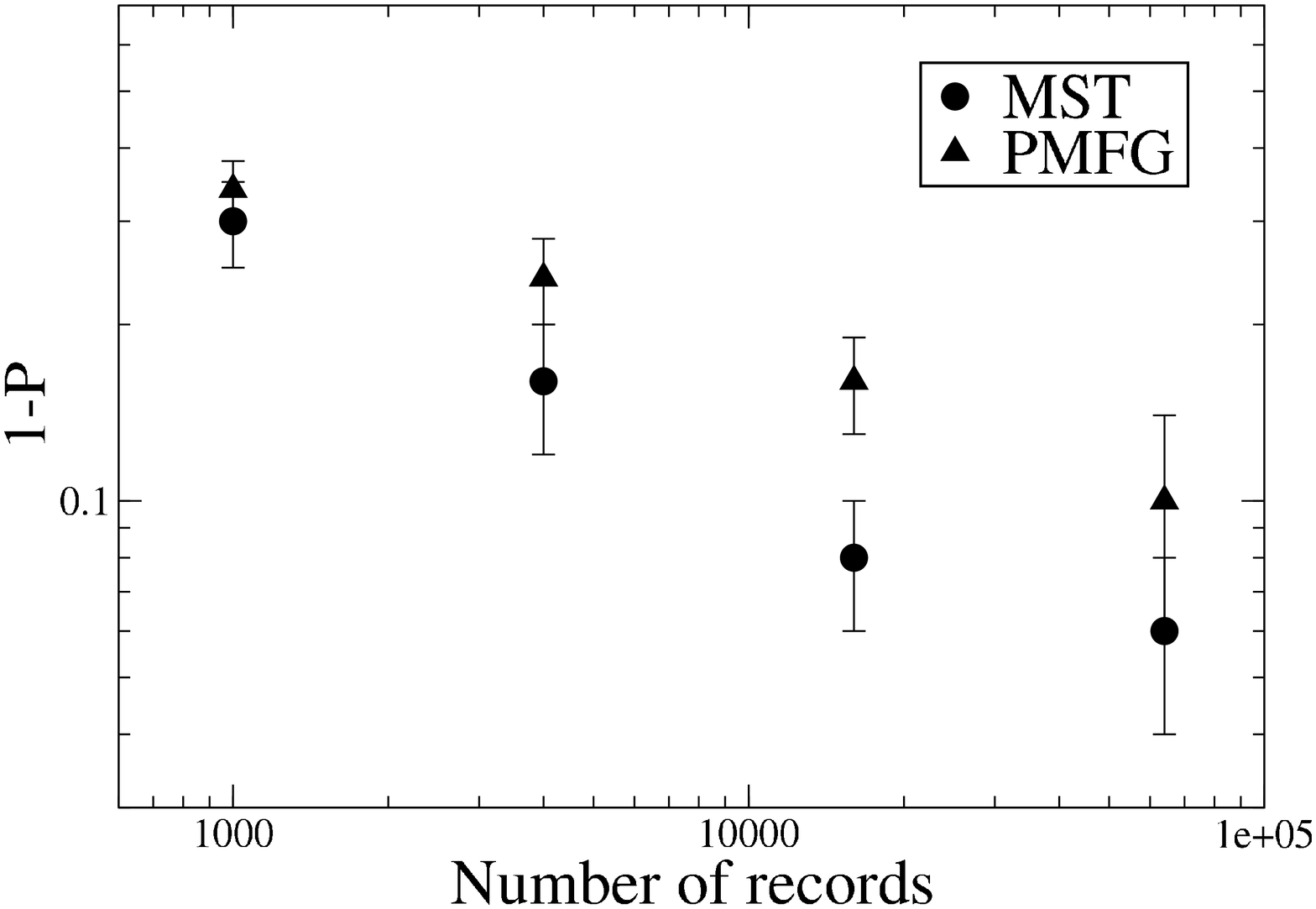}
\caption{\label{stabfig} Analysis of stability of the MST and PMFG with respect to the statistical uncertainty present in the estimation of the correlation matrix as a function of the number of records of the multivariate time series. By assuming as reference matrix the \emph{empirical} correlation matrix of the system we perform 4 sets of 20 realizations each one of surrogated multivariate time series. Each set is characterized by a different number of records set as follows: $1000$, $4000$, $16000$ and $64000$ records. For each one of the simulated realizations, both the MST and PMFG have been constructed. The percent ($1-P$) of the number of links of the simulated graphs non-matching with the links of the MST and PMFG of real data is shown as a function of the number of records of the surrogated time series in a log-log plot. The error bar indicates one standard deviation of $1-P$ computed for each set.} 
\end{figure}

The analyzed correlation structure has a certain degree of statistical uncertainty due to the finite length of time series.
The stability of the filtered graphs with respect to such statistical uncertainty has been analyzed by generating surrogated data  series using the discrete Karhunen-Lo\`eve expansion \cite{johnson94}. The random multivariate Gaussian data series are computed starting from a given correlation matrix.
For any simulated realization the correlation matrix has been calculated. The computed matrices become closer and closer to the reference matrix by increasing the number of records of the simulated time series. We consider the \emph{empirical} correlation matrix associated to the system as the reference matrix. For fixed values of the number of records of time series, $20$ realizations are simulated and for each of them both the MST and the PMFG have been determined.
In Fig. 3 the percent  of non-matching edges in the simulated and in the real data graphs  is plotted as a function of the number of records of the simulated time series in a log-log scale. Fig. 3 shows that the MST is marginally more stable than the PMFG.  Fig. 3  also suggests a power law dependence of the stability of the MST and PMFG with respect to the number of data in the multivariate time series. The significant increase of information gained by the PMFG is therefore fully balancing the marginal decrease of stability for any number of records of the multivariate time series.
This is another reason suggesting that the PMFG and the similar graphs characterized by a low value of the genus are the best compromise allowing to consider a graph richer than the MST but characterized by a similar degree of stability with respect to the statistical uncertainty unavoidably associated with graphs modeling complex systems.\\

%
%

\textbf{Conclusions}\\
In summary we have shown that it is possible to determine a family of graphs having the same hierarchical tree associated to the MST but comprising a larger number of links and allowing closed loops. The amount of filtered information with respect to the one present in the MST increases by increasing the genus. A substantial amount of additional filtered information is already obtained in the case for genus $k=0$ which gives the PMFG.
A correlation-based investigation of a financial portfolio shows that the method is pretty efficient in filtering relevant information about the connection structure both of the whole system and within each cluster. The example presented is representative of a large class of correlation based clustering investigations. For this example, the stability of MST and PMFG with respect to the statistical uncertainty due to the finite length of time series has been investigated suggesting a power law dependence of the stability with respect to the number of records in time series. The stability of MST turned out to be slightly higher than the one of PMFG. The proposed procedure can be applied to a large number of real and correlation based networks when a filtering of the similarity measure matrix is needed.\\

\textbf{Acknowledgments}\\
We wish to thank S.T. Hyde for fruitful discussions and advices.
We acknowledge partial support from research project MIUR 449/97 ``Dinamica di altissima frequenza nei mercati finanziari''. 
MT and RNM thank partial funding support from research projects MIUR-FIRB RBNE01CW3M, NEST-DYSONET 12911 EU project.
TA and TDM acknowledge partial support from ARC Discovery Projects DP0344004 (2003), DP0558183 (2005) and Australian Partnership for Advanced Computing National Facilities (APAC).
%

\end{document}